\documentclass[a4paper,fleqn,usenatbib,useAMS]{mn2e}
\usepackage{graphicx}	
\usepackage[flushleft]{threeparttable}
\usepackage[utf8]{inputenc}
\usepackage{amsmath}	
\usepackage{amssymb}	
\usepackage{multicol}        
\usepackage{bm}		
\usepackage{pdflscape}	
\usepackage{color}
\usepackage{hyperref}
\usepackage{natbib}
\usepackage[T1]{fontenc}
\usepackage{ae,aecompl}
\usepackage{newtxtext,newtxmath}
 
\newcommand\mnras{MNRAS}             
\newcommand\aj{AJ}                   
\newcommand\apjl{ApJ}                
\newcommand\aap{A\&A}                

\title[KiDS-SQuaD]{
KiDS-SQuaD: The KiDS Strongly lensed Quasar Detection project }
\author[C.~Spiniello et al.]{C.~Spiniello$^{1,2}$,  A.~Agnello$^{2}$, N.~R.~Napolitano$^{1}$,   A.~V.~Sergeyev$^{3,4}$,  \and F.~I.~Getman$^{1}$, C.~Tortora$^{5}$, M.~Spavone$^{1}$, M. Bilicki$^{6,7,8}$, H. Buddelmeijer$^{6}$,\and   L.V.E.~Koopmans$^{5}$, K.Kuijken$^{6}$, G.~Vernardos$^{5}$, E.~Bannikova$^{3,4}$, M.~Capaccioli$^{9}$
\\
$^{1}$INAF - Osservatorio Astronomico di Capodimonte, Salita Moiariello, 16, I-80131 Napoli, Italy \\
$^{2}$European Southern Observatory, Karl-Schwarschild-Str. 2, 85748 Garching, Germany \\
$^{3}$Astronomical Institute of Kharkov National University, 61022, 35 Sumskaya St, Kharkov, Ukraine \\
$^{4}$Institute of Radio Astronomy of the National Academy of Sciences of Ukraine \\
$^{5}$Kapteyn Astronomical Institute, University of Groningen, P.O. Box 800, 9700 AV Groningen, the Netherlands \\
$^{6}$Leiden Observatory, Leiden University, P.O.Box 9513, 2300RA Leiden, The Netherlands \\
$^{7}$National Centre for Nuclear Research, Astrophysics Division, P.O. Box 447, PL-90-950 \L \'od\'z, Poland\\
$^{8}$Janusz Gil Institute of Astronomy, University of Zielona G\'ora, ul.
Szafrana 2, 65-516 Zielona G\'{o}ra, Poland \\
$^{9}$ Dip. di Fisica, Universita\' di Napoli Federico II, C.U. di Monte Sant'Angelo, Via Cintia, 80126 Naples, Italy.
}

\date{Last updated 2018 June 1}
\pubyear{2018}
\hypersetup{draft}

\begin{document}
\label{firstpage}
\pagerange{\pageref{firstpage}--\pageref{lastpage}}
\maketitle

\begin{abstract}
New methods have recently been developed to search for strong gravitational lenses, in particular lensed quasars, in wide-field imaging surveys. 
Here, we compare the performance of three different, morphology- and photometry-based methods to find lens candidates within the Kilo-Degree Survey (KiDS) DR3 footprint (440~deg$^2$). The three methods are: i) a multiplet detection in KiDS-DR3 and/or Gaia-DR1, ii) direct modeling of KiDS cutouts and iii) positional offsets between different surveys (KiDS-vs-Gaia, Gaia-vs-2MASS), with purpose-built astrometric recalibrations.
The first benchmark for the methods has been set by the recovery of known lenses. We are able to recover seven out of ten known lenses and pairs of quasars observed in the KiDS DR3 footprint, or eight out of ten with improved selection criteria and looser colour pre-selection.
This success rate reflects the combination of all methods together, which, taken individually, performed significantly worse (four lenses each). One novelty of our analysis is that the comparison of the performances of the different methods has revealed the strengths and weaknesses of the approaches and, most of all, the complementarity. 
We finally provide a list of high-grade candidates found by one or more methods, awaiting spectroscopic follow-up for confirmation. Of these, KiDS~1042+0023 is, to our knowledge, the first confirmed lensed quasar from KiDS, exhibiting two quasar spectra at the same source redshift at either sides of a red galaxy, with uniform flux-ratio $f\approx1.25$ over the wavelength range $0.45\mu\mathrm{m}<\lambda<0.75\mu\mathrm{m}.$
\end{abstract}


\section{Introduction}
\label{intro}
Strongly lensed quasars can provide unique insights into major open issues in cosmology and extragalactic astrophysics, but they are an intrinsically rare class of objects, as they require close alignment of quasars (typically at redshifts $z_{s}\approx 2$ or beyond) with galaxies ($z_{l}\approx0.7$) acting as deflectors or lenses. While the details depend on the population properties of quasars and galaxies, roughly one in $10^4$ quasars is expected to be strongly lensed. 
In particular, \cite{Oguri10} estimated a density of $\sim 0.4$ lensed quasars per square degree. 
 These estimates are further affected by the surveys: the image quality,the observed bands, the depth, and by performance of the techniques used to find these systems.

With the KiDS Strongly lensed QUAsar Detection project, or KiDS-SQuaD, we set out to find as many previously undiscovered gravitational lenses as possible in the Kilo Degree Survey (KiDS), taking advantage of the high quality imaging provided by the survey. 

To reach our aims, we apply different methods to the KiDS data within the current footprint in order to  increase the completeness of our search, which we can estimate by testing our methods blindly on already discovered lenses.  
Our final goal is to increase the number of known lensed quasars and build up a statistically relevant number of lensing systems, spanning a wide range of parameters, such as the mass of the deflector, the redshift and nature of the source, and the lensing geometrical configuration.

In preparation for the final release of KiDS, here we focus on the search for lensed quasars and Nearly Identical Quasar pairs (NIQs) in the third KiDS Data Release (440 deg$^{2}$), adopting different search techniques and comparing their outcomes and performance. Given the exquisite image quality of KiDS, and the bottleneck of colour selection that affected previous searches (e.g., \citealt{Agnello17, Williams17}), here we consider two techniques that mostly exploit morphological information, with colour information being used only at the very first stage of catalogue pre-selection. We also use a third technique that uses optical colour selection directly from KiDS but whose cuts are very loose and complemented with direct image analysis. 

A systematic, and truly blind, comparison of different methods applied to the same dataset has also never been performed before but is necessary.  Every single method can be biased, given its set of assumptions and selection criteria, 
and only by comparing different methods it is possible to understand and quantify these biases, to  maximize the final performance and to find the largest possible number of new lenses.

\hspace{120cm}

The paper is organized as follows. In Section~\ref{kids_intro} we provide a brief introduction to the Kilo Degree Survey and we present the samples we use in the paper. 
In Section~\ref{methods} we introduce and detail each of the search methods highlighting limitations, strength points and possible biases. 
In Section~\ref{recovery} we present a quantitative test on the recovery of already known quasar lenses and NIQs that are covered by the KiDS-DR3 footprint. 
In Section~\ref{results} we discuss our results, including a list of 'high-rank candidates' awaiting spectroscopic confirmation as well as  spectroscopic data of KiDS~1042+0023, which is the first new lensed quasar discovered by our team in KiDS. We also briefly describe some tests we ran on the second Gaia data release to access how and if it would improve our results.   
In Section~\ref{future} we discuss future developments highlighting, in particular, possible alternative QSOs pre-selection criteria or ways to improve our current ones to uncover all the uncovered lenses once KiDS will be completed. 
Finally, we present our final conclusions and a summary of the results of the paper in Section~\ref{conclusions}. 
\section{The Data}
\label{kids_intro}
All strong lensed quasars searches share a similar structure and rely on multi-band survey catalogues as well as image cutouts.  In what follows, we use the same nomenclature already introduced by \citet{Agnello17}, hereafter A17: {\sl objects} are selected at query level from wide-field surveys, {\sl targets} are a sub-sample of objects selected based on their catalogue properties, and {\sl candidates} are a sub-sample of targets further selected based on their images either via visual inspection or cutout modeling. 
Whenever needed, for consistency with previous work, magnitudes are used in their native survey definition, i.e. AB for KiDS and Vega for Gaia, WISE, 2MASS.

For KiDS we always refer to r-band coordinates. 

\subsection{The Kilo Degree Survey} 
The Kilo Degree Survey (KiDS, \citealt{deJong15,deJong17}) is one of the public surveys selected by ESO for the VLT Survey telescope (VST, \citealt{Capaccioli11}), operating from Cerro Paranal (Chile) and equipped with the 1 square degree camera OmegaCAM (\citealt{Kuijken11}).   
Once completed, KiDS will cover 1350 deg$^2$ over $u$, $g$, $r$, and $i$ bands split in two stripes (one in the Northern Galactic cap and one in the Southern cap). 

KiDS has been primarily conceived for weak lensing studies (\citealt{Kuijken15}, \citealt{Hildebrandt17}), given the high spatial resolution of VST ($0.2"$/pixel) and the quite stringent seeing constraints requested for the r-band images (always below $1"$, with mean of $0.7"$). This band is also the deepest observed one with 1800s exposure and a limiting magnitude of 25.1 (5$\sigma$ in $2"$ aperture). 
In all other bands the seeing constraints are less stringent, but sub-arcsec mean seeing is also obtained (\citealt{deJong17}). 

The resulting image quality required for weak lensing turns out to be optimal for other studies like 
surface photometry (Roy et al. 2018, submitted) and search for rare systems like ultra-compact massive galaxies (\citealt{Tortora16}), ultra diffuse galaxies (\citealt{vanderBurg17}) galaxy clusters (\citealt{Radovich17}), and strong gravitational lenses (\citealt{Napolitano16, Petrillo17}).

In this paper we use the multi-band images and aperture-matched source catalogue encompassing all the survey tiles of the first three KiDS data releases (440 deg$^2$, \citealt{deJong17})\footnote{http://kids.strw.leidenuniv.nl/index.php} with the final aim of combining different search methods in order to increase completeness and to find a large number of still undiscovered lensed quasars. 
The work of \citet{Petrillo17} focusses on galaxy-galaxy lensing  using morphological classification method based on a Convolutional Neural Network (CNN) and as such complements our work. 

\subsection{Ancillary Surveys: Gaia-DR1 and WISE} 
To complement the KiDS data, we make use of the Gaia Data Release 1 (DR1) catalogue, consisting of astrometry and photometry for over 1 billion sources brighter than magnitude $G=20.7,$ and the Wide-Infrared Survey Explorer (WISE, \citealt{Wright10}) catalogue. 
The WISE mission provides an all-sky catalogue of astrometry and photometry at 3.4$\mu\mathrm{m}$, 4.6$\mu\mathrm{m}$, 12$\mu\mathrm{m}$ and 22$\mu\mathrm{m}$ mid-infrared bandpasses (hereafter W1, W2, W3 and W4), for 747 million objects, with angular resolution of 6.1, 6.4, 6.5, and 12.0 arcsec respectively for the four bands.

The WISE and Gaia-DR1 catalogues have been queried through the NASA/IPAC Infrared Science Archive\footnote{http://irsa.ipac.caltech.edu/frontpage/}. The cross-matches with Gaia-DR2 (discussed in Section~\ref{results_gaia}) have used the CDS-Xmatch service\footnote{http://cdsxmatch.u-strasbg.fr/xmatch}.

\section{The Methods}
\label{methods}
Our lens search approach is "source-based", which means that it prioritises lensing systems where the multiple images of the source give a larger contribution to the light with than the deflector. 
In particular, we use three methods all based on broad band spectral energy distribution (SED) properties, to isolate quasars-like objects from the tens of millions of extended, extragalactic objects, on the basis of optical and/or infrared colours. 
More specifically, the first two methods are based on the same mid-infrared magnitude/colour pre-selection of quasar-like objects from WISE, and are developed starting from the work presented in A17. 
The third method instead relies on optical (KiDS) $ugr$ colour pre-selection.  


\begin{figure*}
\includegraphics[width=18cm]{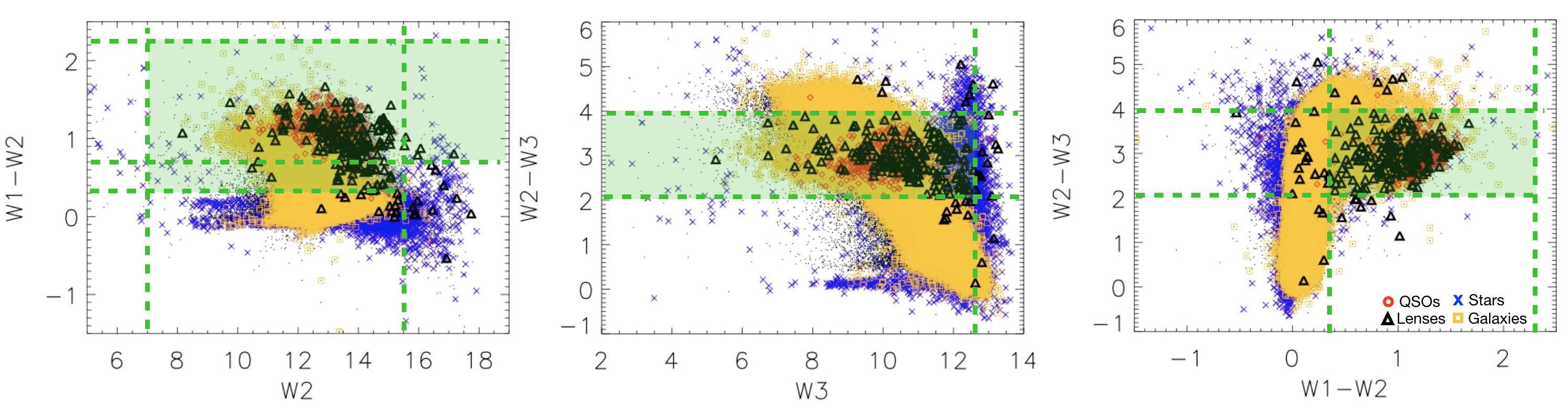}
 \caption{Examples of WISE magnitude- colour-colour plots used to select QSO-like objects. The green lines and boxes show the regions that fulfill the selection criteria of Eq.~1. Points are all the objects from the 6dFGS Survey, with different symbols and colour according to their spectral types: QSO objects are plotted as red circles, stars are plotted as blue crosses and finally galaxies are plotted as yellow squares. Black triangles are all known gravitational QSO lenses and NIQs.}
\label{fig:WISE_colours}
\end{figure*}

After isolating quasars from stars and galaxies with the colour and magnitude cuts, it is necessary to find, among the selected quasars, the very rare ones whose light has been deflected by a galaxy or a cluster and therefore resulted in multiple images. 
To this purpose, we use three approaches which mostly rely on morphological information, with the aim of reducing the colour pre-selection to a minimum, also because the deflector, although not dominant, can still alter the colour of the source quasar.

In the following subsections, we provide a detailed description of each of the methods, explaining their selection criteria and assumptions, in order to quantify their performance and the complementarity of them. 
In fact, one of the goals of this paper is to re-define and improve the selections and procedures in preparation for the final KiDS data release and of future wider/deeper imaging surveys.

\subsection{WISE+KiDS+Gaia Multiplets}
\label{multiplets}
Following A17, we have combined mid-infrared colour selection from WISE and the high spatial resolution of KiDS and Gaia, to identify multiplets of objects with quasar-like colours.  
The use of WISE colors to separate quasars from galaxies and stars, have been already successfully employed by several authors (e.g. \citealt{Stern12, Assef13, Jarrett17}). 
For instance, \cite{Stern12} showed that the W1 - W2 colour, ([3.6] - [4.5] micron) can be used to distinguish quasars from stars based on the fact that the AGN spectrum is generally redder than the blackbody Rayleigh-Jeans stellar spectrum which peaks at rest frame 1.6 $\mu$m. 

We begin with the following cuts in WISE magnitudes and colours to eliminate most stellar contaminants and emission-line galaxies: 
\begin{multline}
7.0 < W2 < 15.6, \\
W3 < 12.6, \\
0.35< (W1 - W2) < 2.3, \\ 
2.1 < (W2 - W3) < 4.0,\\
(W1 - W2) > 0.35 $ \& $ W2 < 14.6 $ OR $ (W1 - W2)>0.7 $ \& $ W2>14.6. 
\end{multline}
These cuts improved the ones used in A17 to isolate QSOs (and even more specifically lensed QSOs) from stars and galaxies. 
In particular, we slightly increased lower and upper limits on magnitudes and colours and inserted a lower limit on W2 and on W1-W2 in order to better match the colour distribution of known lenses (see Fig.~\ref{fig:WISE_colours}). 
To reduce as much as possible the number of selected stellar objects but, at the same time, to try to include as many known lenses as possible, we also used two different thresholds in W1-W2, depending on W2 ($W1-W2>0.35$ for $W2<14.6$, where the contamination from stars is limited, and $W1-W2>0.7$ for $W2>14.6$, where the stellar systems are more broadened in colour). 

The selection criteria are shown as green regions in Figure~\ref{fig:WISE_colours}, where we check our selection against a sample of objects spectroscopically classified by the 6dF Galaxy Survey (6dFGS, \citealt{Jones04, Jones09}) and known lenses.  
In the Figure, data-points are 6dF objects colour-coded by spectral type, whereas the black triangles are already discovered QSO lenses and NIQs in a catalogue of $\approx260$ objects across the entire sky that we assembled from the CASTLES database\footnote{https://www.cfa.harvard.edu/castles/}, The Sloan Digital Sky Survey Quasar Lens Search (SQLS; \citealt{Oguri06}) and recent publications (e.g., \citealt{Sergeyev16,More16, Lemon18}).

Finally, we further restricted our selection to objects with small uncertainties on the WISE magnitudes ($\delta W1<0.25$, $\delta W2<0.3$ and $\delta W3<0.35$). This is the strongest constraint we applied to our object catalogues, but it is necessary to exclude most of the contaminants. 
 
Overall,  the magnitudes and colour cuts are still quite generous, e.g. towards the redder W1-W2 colours or lower values for W3 and W2 magnitudes.
On the other hand,  there is nevertheless room for some improvement since there are still some known quasar lenses (and legitimate quasars in the first place) that escape the cuts applied above. 
A quantitative discussion of recovery of the known lenses is detailed in Sec.\ref{recovery}. In Sec.\ref{future} we will discuss how to optimize the WISE colour/magnitude cuts in order to cover the largest possible number of known lenses, while also limiting contamination. 

After the WISE colour {\sl pre-selection}, we restricted the search first to the KiDS final footprint, which yielded 278994 targets. We then cross-matched this table with Gaia-DR1. We lose at this stage $\sim 72$\% of the selected ojbects but this step is necessary to overcome the large (in same cases up to $\approx 3^{\prime\prime}$) positional errors in the WISE catalogue.

In order to identify KiDS counterparts to the above targets, we matched the WISE-Gaia table with the catalogue of all sources in KiDS-DR3 (\citealt{deJong17}), which covers an area of 440 deg$^2$ in the four $ugri$ bands. This match produced a list of 78377 targets with WISE and optical photometry, Gaia-DR1 coordinates and a counterpart in KiDS.

From the WISE-Gaia-KiDS table we then isolated \textit{multiplets}, i.e. WISE objects that resulted in multiple Gaia or KiDS \texttt{source} entries. Based on previous investigation (e.g. A17), the remaining \textit{singlets} may be truly isolated quasars, or quasars with faint line-of-sight companions, or lenses that are not suitably resolved by the Gaia or KiDS source-extraction pipelines. For this reason, in this search we selected both WISE-Gaia and WISE-KiDS multiplets. 

Our final list comprises 5245 targets for WISE+KiDS and 1277 targets for WISE+Gaia. 
Given the manageable final number, rather than basing our candidate-selection on automatic, colour-based procedures with the risk of losing many interesting objects due to scatter in their population properties, we decided (at least for this pilot program) to visually inspect all targets and assemble a list of graded candidates, which will be presented in Sec.~\ref{candidates}. 
To assign a grade to the candidates, three different members of our team inspected the list of candidates independently, assigning to each of them a grading from 1 (low possibility to be a lens) to 4 (sure lens or known lens). The final grade for a given system is then the average of the three grades. 
 The same grading procedure has been repeated also for the other methods.

There are three main limitations of this first method. First, in order to have multiple matches in KiDS or Gaia, the different components must be deblended by the Source Extractor routine used to create the source catalogues. Second, we set up a maximum separation between the components of $5"$, motivated by the average separation of known lenses and by the WISE image quality. 
In this way, we might lose some lenses with a wider separation, but we note that these are extremely rare cases. 
Third, we required that the objects satisfying the WISE selection criteria are bright enough to have a Gaia counterpart.  

The final list of candidates accepted after visual inspection comprises $\approx20$ very promising candidates, with grade $\ge1$.

\subsection{BaROQuES: Blue and Red Offsets of Quasars and Extragalactic sources}
\label{baroques}
If the deflector and quasar images contribute differently in different bands, this should result in centroid offsets of the same object among different surveys. 
While this morphological selection was proposed in the past (e.g. \citealt{Agnello15a}), the astrometric accuracy of ground-based survey catalogues prevented any practical application. 
This changed with the advent of Gaia: exploiting a cross-match of Gaia-DR1 and SDSS (see \citealt{Deason17}), \citet{Lemon17} examined astrometric offsets of quasars in order to isolate lens candidates. However, the astrometric solutions depend on the astrometric standards that are adopted (typically, bright stars). 
Thus, quasars and other extragalactic objects with different spectral energy distributions (SEDs) will have high offsets even if they are not lensed, as a consequence of atmospheric differential refraction (ADR). Reportedly, significant contamination by isolated quasars 
is present in the search by \citet{Lemon17}. In order to resolve this, we concentrate exclusively on extra-galactic sources (in particular, photometrically selected quasars) and re-compute a local astrometric solution based directly on these.
Despite the variety of SEDs exhibited by WISE-selected quasars, this choice is sufficient to mitigate the spurious offsets induced by ADR.

Identical to the previous method, we pre-selected QSO-like objects based on WISE cuts and we cross-matched this table with the Gaia-DR1 and KiDS catalogues. 
We therefore start from the previously obtained table with $78377$ targets satisfying the WISE color cuts, having a counterpart in Gaia-DR1 and covered by the KiDS DR3 footprint. 
Then, we used our own scripts: the Blue and Red Offsets of Quasars and Extragalactic Sources (BaROQuES) scripts\footnote{available upon request, \texttt{https://github.com/aagnello}}
to compute detrended offsets between the positions of WISE-selected objects among different surveys. In order to compute field-corrected offsets, we either used KiDS versus Gaia, or 2MASS versus Gaia.
For each object in a survey cross-match, we considered all neighbouring objects within a chosen window (a radius of 0.5 deg for KiDS-Gaia and 1.0 deg for 2MASS-Gaia), such that $\sim 40$ objects per patch are present, ensuring robust statistics\footnote{We emphasize that, even though nominally there are more than $40$ Gaia-KiDS objects per square degree, in our procedure we are considering only WISE-selected quasar-like objects. The choice of bounding patch (i.e. as a circle, or as a square in $\delta\mathrm{R.A.}$ and $\delta\mathrm{Dec.}$, or as a square in $\cos(\mathrm{Dec.})\delta\mathrm{R.A.}$ and $\delta\mathrm{Dec.}$) does not have any appreciable effect.}.

We then computed the average offset in right ascension and declination over that patch, and subtracted it from the offset computed on the considered object. We then retained just those objects that have an average-subtracted offset greater than a prescribed threshold (0.3$^{\prime\prime}$ for 2MASS-Gaia, 0.2$^{\prime\prime}$ for KiDS-Gaia). The choice of threshold is mainly dictated by the accuracy of KiDS/2MASS, and is such that most (if not all) of known lenses are still recovered after the cuts in field-corrected offsets.

The final sample of KiDS-Gaia `baroques' comprises 8180 targets (10\% of the original table), of which only 594 (7\% of the targets) are also multiplets.
Also in this case,  we proceeded with visual inspection and grading to build the final list of candidates. Contamination by isolated quasars is null.
This zeroth-order, local astrometric solution is already sufficient to cull isolated quasars, even though in general 
 astrometric solutions can be non-linear on degree-wide scales. 

The main limitation of this second method is that it mainly selects systems where both lens and source give a non-negligible contribution to the light, whereas it would most probably lose lenses where one of the components dominates.   
On the contrary, the Multiplets method is in principle able to find pairs where the lens is not visible (or not there at all, i.e. NIQs) and also objects where the lens is more dominant, as long as the QSOs are deblended and resolved as multiple objects in the KiDS/Gaia catalogue. 
This is why the combination of these two WISE+KiDS methods is important towards maximizing completeness.

After visual inspection, the BaROQuES method produced $40$ candidates with grade $>1$, of which 5 in common with the Multiplet method (see Table~\ref{tab:candidates_best}).

\begin{table*}
\caption{Ten known lensed quasars observed by KiDS DR3. For each lens, we provide the methods that recover it in Col.4 and the reasons why it was excluded from the selection of a give method in the footnotes (indentified by numbers). }
\label{tab:known_in_kids}
\begin{center}
\begin{tabular}{l | c | c | c | c }
\hline
{\bf ID}  &	 {\bf RA (J2000)} &	 {\bf DEC (J2000)}   & {\bf Found with (Lost because -see footnotes)} & {\bf Reference}\\
 \hline
SDSS1226-0006		&		$12$:$26$:$08.16$  &  $-00$:$06$:$02.02$  	& Multiplets, BaROQuES, DIA	 &	Pindor et al. (2003)	\\
SDSS0924+0219		&		$09$:$24$:$55.92$  & $+02$:$19$:$24.89$   & Multiplets, BaROQuES, DIA   & Inada et al.(2003)			\\	
A0326-3122 (NIQ)			&		$03$:$26$:$06.79$  &  $-31$:$22$:$53.76$	& Multiplets, BaROQuES (1) & Schechter et al. (2017)\\
LBQS1429-008 (NIQ)		&		$14$:$32$:$29.04$  &  $-01$:$06$:$13.00$   & BaROQuES (1,2)  &	Hewett et al. (1989)		\\	
QJ0240-343 (NIQ)			&		$02$:$40$:$07.70$  &  $-34$:$34$:$19.92$	& Multiplets (1,4) &	 Tinney et al. (1995)	\\
WISE2344-3056		&		$23$:$44$:$17.04$  &  $-30$:$56$:$26.52$   & DIA (3, 5) & Schechter et al. (2017)\\
HSC115252+004733 &		$11$:$52$:$52.25$ &	$+00$:$47$:$33.00$ & DIA (6)	& More  et al. (2017) \\
\hline
2QZJ1427-0121A	&		$14$:$27$:$58.80$  &  $-01$:$21$:$31.00$   & (7) &	Hennawi et al.(2006)\\
\hline
SDSS1458-0202	&		$14$:$58$:$47.52$  &  $-02$:$02$:$05.89$ & (8)	& More et al. (2015)\\
CY2201-3201			&		$22$:$01$:$32.88$  & $-32$:$01$:$44.04$  & (8) 	& Castander et al. (2006)		\\
\hline
\end{tabular}
\begin{tablenotes}
\item {{\sl 1}: Lost in DIA: deflector too faint to be detected or not there at all (NIQ).} 
\item {{\sl 2}: Lost in Multiplet: separation bigger than 5\arcsec. } 
\item {{\sl 3}: Lost in Multiplet: not deblended because very low separation between components.}
\item {{\sl 4}: Lost in BaROQuES: deflectors does not give detectable contribution to the light, thus it does not create a shift in positions.} 
\item {{\sl 5}: Lost in BaROQuES and/or Multiplet: missing counterpart in Gaia-DR1.}
\item {{\sl 6}: Lost because of WISE colour selection:  deflector dominates the light.}
\item{{\sl 7}: Lost because of WISE colour selection, but will be recovered with new criteria presented in Sec.~\ref{future}.}
\item{{\sl 8}: Lost because of WISE colour selection: no error is given on W3 and W4,i.e.the source is an upper limit or non-detection in the WISE database.}
  \end{tablenotes}
\end{center}
\end{table*}

\subsection{Direct Image Analysis (DIA)}
\label{dia}
This method is based on direct analysis of KiDS images, without any cross-match with other surveys. A pre-selection of QSO-like candidates is also performed, but based on optical colours, as we will detail later. 

We begin with considering all the objects extracted from the u-band catalog (i.e., $4915119$). 
For each of them we look for multiple detections within a circular aperture radius of $5\arcsec$ in the r-band images, since they have the best depth and resolution, finding 104045 matches. 

As second step, we separate the systems into point-like and extended sources.  We estimate from the r-bands a “point-like-magnitude" for each object under the assumption of a Gaussian distribution, centered on the pixel with maximum flux and matching the full-width half maximum (FWHM) of the image.  
We then compared this "Gaussianized-magnitude" with the aperture magnitude obtained within the FWHM and target the object as "extended" when the two values differ by more than one magnitude with respect to the locus of the sure stars in the same image (\citealt{deJong17}).

At this point, we calculated the magnitudes in $ugr$-bands and the $(u-g)$ and $(g-r)$ colours for all the multiple point-like objects and we used them to select QSO-like candidates according to the following cuts: $(u-g)<1.5$ and $(g-r)<0.7$.  
We calculated that these cuts select 89\% of the QSOs in the Sloan Digital Sky Survey IV (SDSS, \citealt{Blanton17}). Despite a slight difference in the SDSS and KiDS colours (0.05 mag), we decided to adopt the same criteria for our dataset, at least in this pilot program. 
A shortcoming of these colour criteria is that we might lose systems with redshifts $z > 3$, which have $(g-r) > 1$. 
The limitations of these assumptions and different ways to improve our selection criteria will be discussed in Sec.\ref{future}. 

The $12396$ candidates of the colour-selected list () are finally processed with a semi-automatic PSF-subtraction procedure to check for the presence of the deflector (residuals between the multiple subtracted components).
For this task we used the PyRAF package \footnote{PyRAF is a product of the Space Telescope Science Institute, which is operated by AURA for NASA.}.
In particular, from the r-band images we detected all sources within $10"$ radius from the central object and we fit a PSF model to them. 
We generated subtracted images which we visually inspected to check for the presence of residuals from the deflector. 
For completeness we also visually inspected the corresponding combined colour ($ugr$) images to check that all of the QSO multiple images have similar colours, that the geometrical configuration is compatible with a strong lensing configuration and that there are no starburst areas or star forming regions near by.
Finally, {as additional sanity check,} we extended the visual inspection also to the sources that were initially tagged as “extended”  to ensure that we did not mis-classify quadruplets and low-separation multiple systems {that were not deblended in the KiDS catalog. We did not find any mis-classified object.}

The final list of candidates accepted after visual inspection comprises $\approx30$ very promising candidates, with only six already selected by our previous methods.

\section{Recovery of known lenses}
\label{recovery}


The zeroth-order benchmark of the methods discussed in the previous paragraphs is given by the fraction of the known lensed quasars and NIQs that we have "blindly" recovered by our techniques\footnote{We build up a list of candidates for each method and only then, a posteriori, we cross-match it with a catalogue of known QSO lenses identified the ones, among the candidates, that have already been confirmed as lenses.}. 

Among the $\approx260$ confirmed lenses, ten are in the KiDS-DR3. 
In Table~\ref{tab:known_in_kids} we report their coordinates and their discovery papers.
Of course, the full assessment of the overall true selection process efficiency and of the purity of our methods will come only with a significant spectroscopic sample, which we have started to assemble as detailed in Sec.~\ref{results}.

As already anticipated, looking at the missed lenses, the main failure point is the WISE pre-selection.
With the selection criteria of Eq.~1, four lenses would have been lost in the colour pre-selection on the basis of their magnitudes and colours. 
One of them (HSC1152+0047), however, has been found by the DIA method, which is not based on WISE pre-selection.

We repeated the test for the 22 known lenses that will be covered by the final KiDS footprint finding that seven of them will not pass the pre-selection criteria, mainly 
because of their W1-W2 colour. 
This could be caused by the fact that, in case the lens and the source are blended in WISE and the deflector gives a large contribution to the light, the colours of this {\sl effective} source may be not quasar-like anymore and move indeed toward lower W1-W2 values. 
Thus, allowing for a lower threshold in W1-W2 would help in selecting some of the lost known lenses, including one already observed in KiDS-DR3 (2QZJ1427-0121A). 

Finally, we run an analogous test for the entire catalog of known lenses,showed as black triangles in Figure~\ref{fig:WISE_colours}, to increase the number statistics and above all to assess the purity/completeness trade-off in our WISE pre-selection.  Roughly one fourth (40 out of 260) of all the known lensed quasars and NIQs do not pass the WISE pre-selection. This is due primarily to missing entries for the WISE magnitude errors, which are part of the WISE-Gaia pre-selection. 

We conclude that our selection criteria must be optimized if we aim at 
find a large number of new gravitational lenses covered by KiDS, once completed, and by the DR2 of Gaia.  
We will discuss more flexible WISE pre-selections or possible alternatives to it in Sec.~\ref{future}.



As highlighted in Table~\ref{tab:known_in_kids}, seven out of ten known lenses have been found by at least one of our methods, and only two have been recovered by all of them. 
Singularly, the three methods performed equally well, recovering four lenses each,  
and they resulted very complementary since the combination of them bring completeness from 40\% to 70\%. 

\begin{figure*}
\includegraphics[width=17cm]{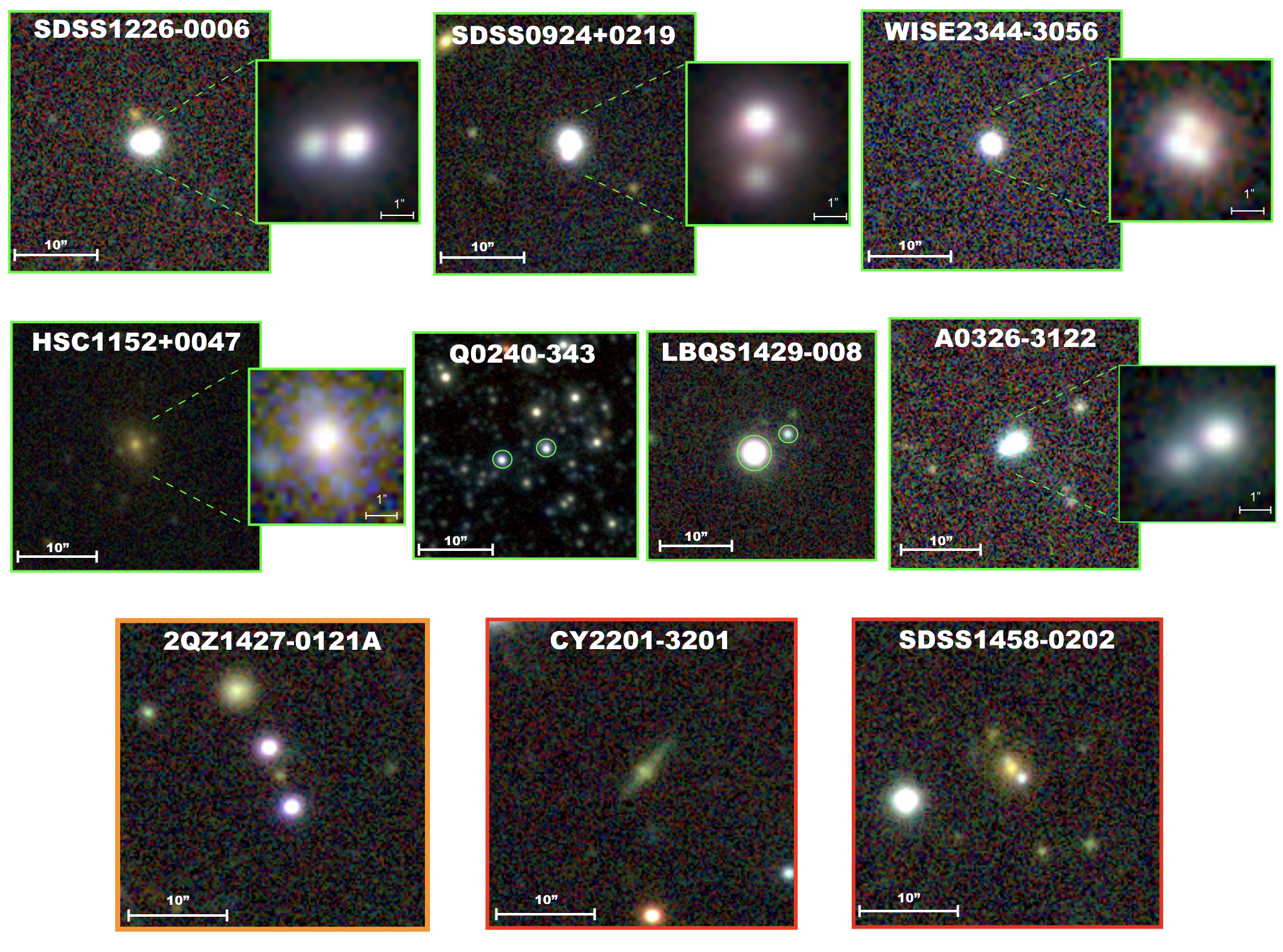}
 \caption{KiDS colour cutouts (g, r, i) of known lenses. $25" \times 25"$  cutouts of KiDS images of the known gravitational lenses already observed by KiDS-DR3. For the systems with low separation between components, we also show a zoom-in of $5" \times 5"$ centered on the system. 
 The first two rows show the known lenses that we found with at least one of our method. The last row shows instead the three lenses that we did not recover. The leftmost lens (orange box) would have been recovered with the new criteria on WISE magnitude and colours presented in Sec.~\ref{future}}. 
\label{fig:known}
\end{figure*}

In the following, we examine each known, recovered lens, highlighting by which method it was found/missed and providing an explanation. 
Finally, in Figure~\ref{fig:known} we show $25" \times 25"$  KiDS cutouts (colour combination of g, r, and i bands). For some lenses, we also show a zoom-in of $5" \times 5"$ to highlight the exquisite image quality of KiDS that enables a clear recognition of multiple images and, in most cases, also the deflector.

\subsection{The easiest cases: SDSS1226-0006 and SDSS0924+0219} 
SDSS1226-0006 is a doublet discovered by \citet{Pindor03} and with a separation within the two QSO images of $\approx1.2"$,  whereas SDSS0924+0219 is a quadruple system, discovered by \citet{Inada03}, with a maximum separation between components of $\approx1.8"$. 
Both systems have a separation large enough to be deblended into multiple components by the KiDS catalogue and with a deflector bright enough to cause an offset in the center positions between KiDS and Gaia. 
We therefore easily recovered these two candidates with all our methods. 


\subsection{ Overcoming the catalogs limitations: WISE2344-3056 and HSC1152+0047}
More difficult is the case of WISE 2344-3056, recently discovered in the VLT-ATLAS Survey by \cite{Schechter17}, which has a very small separation between the multiple images and thus it has not been classified as a multiplet in KiDS. 
Moreover, unfortunately this quad does not have a match in Gaia-DR1 and therefore we could not recover it with the BaROQuES method.  The only method that could, and did, recover this lens was the DIA and it succeeded in finding it. 

Another "difficult case" is HSC1152+0047, the quadruplet serendipitously discovered in the Hyper Suprime-Cam (HSC) Survey (\citealt{More17}). 
The source at $z_{s}=3.76$ is lensed by an early-type galaxy at $z_{l}=0.466$ and a satellite galaxy and it is unusually compact and faint. This causes the deflector to dominate the WISE colours and therefore the exclusion from our colour and magnitude pre-selection criteria.

In this two last cases, the lenses have been recovered only by DIA, which is not based on WISE nor on catalogue cross-matching.  However, in other cases, as the ones described in the next section DIA is not able to recover known lenses found instead by the BaROQuES or the Multiplet methods. 
Thus, we wish to highlight the complementarity of our methods, based on different assumptions and techniques.

\begin{table*}
\caption{High-grade candidates from KiDS-DR3 that have been found by more than one method  
A list of high-grade candidates found by a single method is provided in Table~\ref{tab:candidates_all}.}
\label{tab:candidates_best}
\begin{center}
\begin{tabular}{lccccl}
\hline
{\bf ID}		&	{\bf RA (J2000)}	&	{\bf DEC (J2000)}	&	{\bf Methods} 	& 	{\bf Grade}& 	{\bf Notes} \\
\hline
KIDS0848+0115	& 	08$:$48$:$56	& 	+01$:$15$:$39 	& 	Multipl., BaROQuES, DIA & 2.5 & One of the images has a QSO SDSS spectrum (z$=0.645$) \\	
KIDS2307-3039	& 	23$:$07$:$18	& 	-30$:$39$:$15 	& 	Multipl., BaROQuES, DIA & 2.5 &	  \\
KIDS0841+0101	& 	08$:$41$:$35 & 	+01$:$01$:$56 	& 	Multipl., 	BaROQuES	& 2.5 		& Possible gravitational arc \\						
KIDS1217-0256	& 	12$:$17$:$09	& 	-02$:$56$:$21 		& 	Multipl., BaROQuES	& 2.5		& 	\\ 
KIDS2316-3320	& 	23$:$16$:$27	& 	-33$:$20$:$02 		& 	Multipl., BaROQuES	& 2.5	 	&	Possible NIQ \\	
KIDS0324-3042	&	03$:$24$:$27 	&	-30$:$42$:$50 & 	Multipl., DIA 	& 2.5	& \\
KIDS0924-0128	& 	09$:$24$:$37	& 	-01$:$28$:$44 	& 	Multipl., DIA	&  3.0	& One of the images has a QSO SDSS spectrum (z$=2.446$) \\
KIDS1441+0237	& 	14$:$41$:$45	& 	+02$:$37$:$43 	& 	Multipl., DIA	&  3.0 & One of the images has a QSO SDSS spectrum (z$=1.61$) \\
KIDS1042+0023	&	10$:$42$:$37	& 	+00$:$23$:$02 	& 	Multipl., DIA 	& 3.5	& Spectroscopic data confirmed the lensing nature (Fig.~\ref{fig:spectra1042}) \\
\hline		
\hline
\end{tabular}
  \end{center}
\end{table*}

\subsection{Nearly Identical Quasar pairs: A0326-3122, QJ0240-343 and LBQS1429-008}
NIQs are pairs of quasars exhibiting the same lines at the same redshift and uniform flux-ratios with wavelength. These could be multiple images of the same object deflected by a very faint detector or truly physical pairs. 
In the case of nearly identical quasar pairs the DIA method is not ideal.  Given that the deflector is either too faint to be detected or simply not there, these systems would be automatic discarded when  analyzing the PSF subtracted images (since no residuals would be identified) although they might passed the colours pre-selection and be  triggered by the automatic pipeline. 

A0326-3122 was found by \cite{Schechter17} in the VLT-ATLAS DR2 survey, 
QJ0240-343 was identified by \cite{Tinney95} as a $z=1.406$ candidate gravitational lens system behind the Fornax dwarf spheroidal galaxy, and LBQS1429-008 is a wide-separation binary quasar ($\approx5"$) discovered by \cite{Hewett89}.  

The first two NIQs have been found by our Multiplets method. 
LBQS1429-008 has a separation between the components larger than $5"$, which is the limit we set for Multiplet and DIA. 
This system and A0326-3122 have been found by the BaROQuES.  This might indicate the presence of a deflector, which is however very faint and therefore not visible from the KiDS images. Deeper spectroscopic follow-up or higher-resolution imaging are necessary to assess the lensing nature of these systems.


\section{Results}
\label{results}
The KiDS Strongly lensed QUAsar Detection project (KiDS-SQuaD) aims at finding strong gravitational lensing events in the Kilo Degree Survey (KiDS).
Here, we focus on finding  lensed quasars with the goal of optimizing the filter criteria in a combination of infrared colours from WISE and optical colours from KiDS to get to the closest number of confirmed quasars to the expected one. 

\subsection{High-grade targets, and the first spectroscopically confirmed lens}
\label{candidates}
To further demonstrate that the methods are producing valuable results, we provide in Table~\ref{tab:candidates_best} a list of high-graded candidates in KiDS-DR3 found with more than one method (grade$>2.5$, with a maximum of 4 for known lenses). A more complete list of candidates with grade$\geq 1$ found also by only one of the methods is given in Table~\ref{tab:candidates_all} and 4. 
%
Three of the most promising candidates are also shown in Figure~3.

\begin{figure}
\includegraphics[width=\columnwidth]{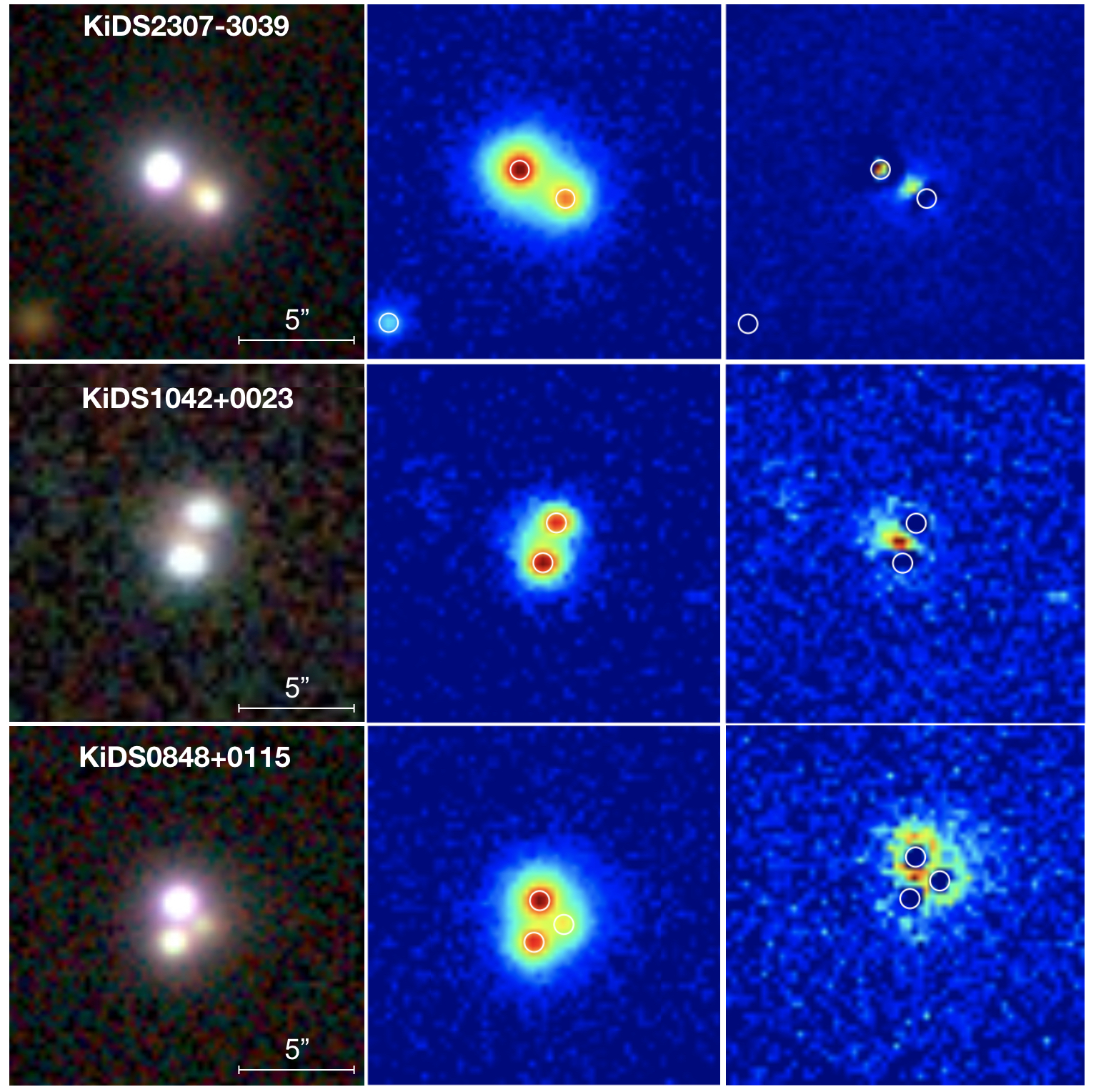}
\label{fig:bestcandidates_new}
\caption{Three of the most promising candidates found by our methods. The left panel of each row shows the g-r-i combined image from KiDS, the middle panel shows the KiDS image in r-band only and the rightmost panel shows the same image after PSF subtraction of the QSO images, revealing the presence of the deflector. }
\end{figure}

A spectroscopic confirmation of the targets selected with the combination of the three methods is beyond the purpose of this first paper. A dedicated, spectroscopic campaigns over a sizeable candidate sample is already planned for future papers of the series and will enable us to quantify the performance of our multi-method strategy.
Nevertheless, we were able to obtain long-slit spectra from the Telescopio Nazionale Galileo (TNG) for the candidates KiDS1042+0023 and KIDS0834-0139. 
We observed both systems with DOLORES (Device Optimized for the LOw RESolution) in long-slit mode (slit width of $1.5"$) using the LR-B grism (wavelength range: $0.3\mu\mathrm{m}<\lambda<0.843\mu\mathrm{m}$, dispersion: $2.52$ \AA/px) for a total integration time per target of 1200 seconds.

\begin{figure}
\includegraphics[width=\columnwidth]{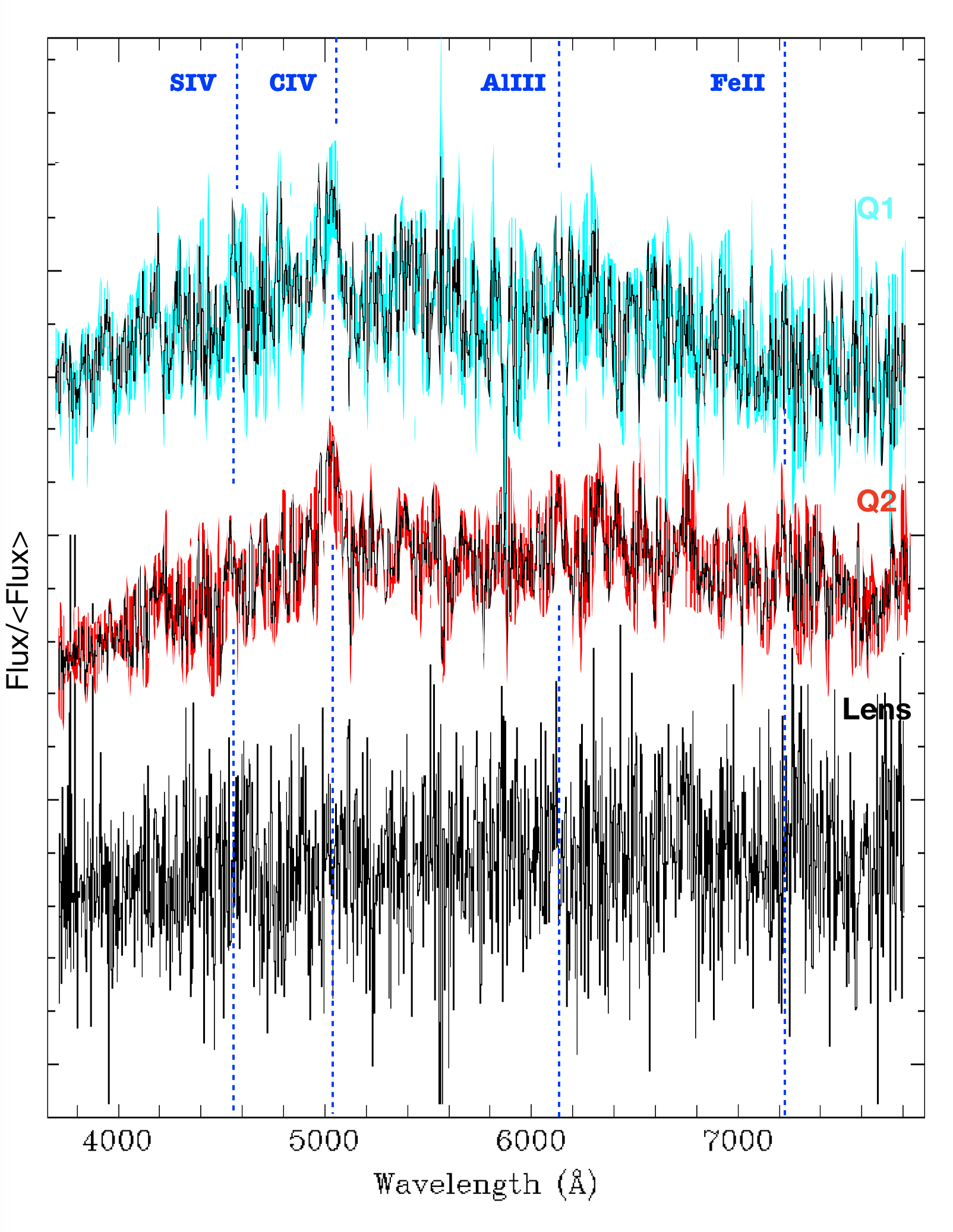}
\caption{TNG spectra of KIDS1042. The light blue and red spectra at the top represent the QSO traces while the black spectrum on the bottom is extracted at the position of the deflector. The spectra have been normalized and plotted with an ad-hoc shift for visualization purposes. For the two QSO spectra, a smoother version (boxcar of 3 pixels) is also overplotted for clarity. The blue dotted vertical lines highlight possible emission lines, corresponding to $z\approx2.26$ for the source.}
\label{fig:spectra1042}
\end{figure}

The spectra of KiDS1042+0023(R.A.=10:42:37.255, Dec.=00:23:02.064), although of very low signal-to-noise (S/N$\approx3-5$ per pixel) show features typical of lensed quasars, confirming the very first lensed QSO discovered in the KiDS survey. We extracted a 1D spectrum at the position of the two multiple images as well as a spectrum in the middle, where the lens lies\footnote{The separation of traces, based on direct modeling of the spatial profiles, fully accounts for cross-contamination and performs well even in very unfavourable seeing conditions (see e.g. \citealt{Agnello18}).}. 
Despite the very poor S/N, the presence of two quasar spectra at the same source redshift is clear, with uniform flux-ratio $f\approx1.25$ over $0.45\mu\mathrm{m}<\lambda<0.75\mu\mathrm{m}.$
The spectra of quasar images (red and light blue with a smoothed version superimposed as black lines) and lens (black spectrum) are shown in Figure~\ref{fig:spectra1042}, in units of normalized, not calibrated flux, with some possible emission lines highlighted with blue vertical lines corresponding to $z\approx2.26$. 
The two quasars spectra have been vertically shifted for visualization purposes. 

Unfortunately, instead, KIDS0834-0139 turned out to be a line-of-sight pair of two white dwarfs stars. 

\subsection{Spectroscopic Surveys and KiDS}
All the methods described here were based on colour pre-selection of point-like, extragalactic objects (through infrared colours from WISE or optical from KiDS). 
An alternative approach is to select these objects on the basis of their spectra. 
To test this, we extracted all the objects from the 6dFGS with spectroscopic redshift $z>0.5$ and good quality flag (quality $\ge3$, i.e. all the red points in Fig.~\ref{fig:WISE_colours}). 
Then, we cross-matched them with the KiDS catalogue and proceeded to visually inspecting the 400 resulting objects. We found three objects already found by other methods and no other very good candidates. 
We repeated the experiment using the 2dF QSO Redshift Survey (2QZ, \citealt{Boyle00}), finding 8164 targets, classified as QSOs, and, after visual inspection, again no "new" promising (grade> 2) candidates.  
Further tests with other spectroscopic surveys (e.g. GAMA, \citealt{Driver08, Driver11}) will be performed in future papers. 

\subsection{Gaia-DR2}
\label{results_gaia}
The recent second Data Release of Gaia has made colours and proper motions available, and should have a nominal resolution of $0.4^{\prime\prime}.$ 
To check the improvements with respect to Gaia DR1 and to exploit the possibility of using proper motion to separate stars from quasars, we ran a cross-match query between WISE, Gaia-DR1 and Gaia DR2. 

We imposed very generous constraints on WISE magnitudes and colors, in order to exclude stars without penalizing extra-galactic objects. To this purpose, we also updated the selection criteria on the WISE errors, using the interpolating formulae given in Eq.~3-4, in order to circumvent missing entries.

We took the coordinates of our singlet BaROQuES candidates (in KiDS-Gaia DR1), and queried Gaia-DR2 on these to check how many objects which were not resolved as "multiplets" in DR1 are now with the new Gaia release. A small fraction (3-5\%) of the DR1 `baroque singlets' are recognized as multiplets by DR2.  
A puzzling outcome is that some Gaia-DR1 sources, bright enough to have perfectly trustworthy 2MASS magnitudes, have now disappeared in Gaia-DR2.

%
%


Moreover, for completeness, we also ran some tests on recovery of known lenses for Gaia-DR2. 
Of 260 known lenses/pairs in our list, 43 are missing from DR2; of the 171 that are detected, 124 are resolved in two sources, 15 into three sources, and six into four sources. 
An interesting feature is that known lenses can have significant proper motions (as determined by the Gaia pipeline). This is valid also for the DR1 baroque singlets (lens candidates), and prevents further separation between lenses and contaminants (e.g. line-of-sight alignments of quasars and stars) based on Gaia-DR2 proper motions alone. 

Thus, for the purpose of this paper, we conclude that proper motions do not help considerably, at least not when BaROQuES singlets are concerned. What seems to be marginally helpful is a combination of Gaia colours
\begin{equation}
\mathrm{gcol} = G-G_{RP}-(G_{BP}-G)/0.75 \ ,
\end{equation}
which seems to isolate white dwarfs (WDs) and possibly galaxies from a more reliable sample of quasars. 
In fact, almost all of the known lenses/pairs in our produced table, except a handful, have gcol>0.15. However, the reader should keep in mind that this criterion requires Gaia-DR2 colours, which is not always the case even on known lenses. 

\section{Future Prospects}
\label{future}
In order to further improve the lens search, it will be certainly necessary to work on the WISE pre-selection cuts, al largely shown in this paper. 
To this purpose, we highlight four possibilities.
\begin{enumerate}
\item Looser pre-selections cuts. \\
\noindent Magnitude errors are sometimes missing from the WISE catalogues. One might think of relaxing the cuts completely removing the constraints on the uncertainties, but in this case, the number of targets would be too large for visual inspection (more than seven millions of objects at pre-selection stage), and would include an overwhelming fraction of galactic and extragalactic contaminants.
We notice that when errors are available they are well fitted by 
\begin{eqnarray}
\nonumber \delta W1 = 0.024\times \sqrt{1.0 + \exp{[1.5\times(W1-15.0)]}}, \\
\nonumber \delta W2 = 0.021\times\sqrt{1.0 + \exp{[1.5\times(W2-13.3)]}}, \\
\delta W3 = 0.016\times \sqrt{1.0 + \exp{[1.7\times(W3-8.6)]. }}
\end{eqnarray}
This partially relieves the issues with criteria relying on magnitude errors. 
Also smoother cuts in WISE magnitudes can be used in order to exclude stars at some prescribed confidence level. From the fitting formulae above, a simple (and currently used) alternative to the original WISE preselection is
\begin{eqnarray}
\nonumber W1-W2>0.2+\sqrt{(\delta W1)^{2}+(\delta W2)^{2}},\\
W2-W3>2.0+\sqrt{(\delta W2)^{2}+(\delta W3)^{2}};
\end{eqnarray}
once tested on the 6dFGS catalogue, this criterion eliminates most of the stars and retains most of the extragalactic sources. 
Despite the improvement, however, some systems with $W1-W2\approx W2-W3\approx 0$ (e.g. CY2201-3201) are still lost at pre-selection.

\item Estimation of magnitudes from WISE fluxes. \\
\noindent The cases when an error estimate is not provided correspond to either a low S/N ($<2$) in the given band (mostly) or artifacts.  For S/N$<2$ sources, the WISE public database reports upper limits in magnitudes which are significantly biased. 
A possible way of getting the proper magnitude instead of the upper limit
for sources with low S/N, is by using the flux from the WISE database, translating them into magnitudes. 

\item Population-mixture classification \citep[e.g.][]{Williams17}, which can be used to exclude the most abundant contaminant classes combined with a looser WISE pre-selection without any constraints on the errors on the magnitudes. 

\item Forced-photometry. \\
\noindent WISE catalogue magnitudes become quickly unreliable beyond W1=17.0, W2=15.6, W3=11.8, where $\approx20\%$ of known lenses lie. 
Furthermore, we note that  if one uses the public WISE database, then the source selection will be limited. This is because the IPAC database requires
that each source listed has a 5$\sigma$ detection (S/N$\geq5$) in at
least one of the four WISE bands. This means that hypothetical sources
having a detection in {\sl all} thee bands only slightly lower than the threshold  
will be discarded.  This is a major limitation for surveys such as KiDS, whose 
depth and image quality would justify a forced-photometry effort (e.g. \citealt{Lang16}), which will increase the number of detections.   
\end{enumerate}


The current optical colour pre-selection (used in the DIA) mostly loses high redshift QSOs ($z>3,$ which typically have $(g-r)>1.5$). However, including these objects will also include a very large contamination from extended sources such as galaxies. Two possibilities for improvement, for optical-based pre-selection, would be a variability criterion and the use of machine learning techniques for quasar selection (some of which are already available, e.g. {\citealt{Chehade16}).


Furthermore, we note that the KiDS Survey is twinned with the NIR survey VIKING (Edge et al. 2014) from VISTA/VIRCAM also located in Cerro Paranal. VIKING covers the same area of KiDS in ZYJHK,  making the multi-band dataset from the two surveys ideal to perform object selection based on a wide baseline that would efficiently pinpoint high redshift systems. 
Thus,  KiDS+VIKING, together, provide a unique sensitive, 9-band multi-colour survey, which will lead to the detection of a statistically significant number of new strong lenses. 

Finally, spectra of most Gaia sources will be available upon mission completion (2020). In principle, this will allow for a complete by-passing of colour-magnitude cuts, provided quasars are suitably recognized by the Gaia classification pipelines.



\section{Summary and conclusions}
\label{conclusions}
In this first paper of the the KiDS Strongly lensed QUAsar Detection Project we have performed a truly blind comparison of different methods to find lensed quasars on an homogeneous dataset. 
We have used the state-of-art-techniques based on mid-infrared photometry from WISE, cross-matched with wide surveys (e.g. Gaia, 2MASS), optical colour pre-selection and PSF-analysis, all of them followed by visual inspection of high resolution multi-band KiDS images. 
We have applied all these methods to the KiDS DR3 footprint, with the goal of overtaking the limitation of each method alone and increasing the efficiency of our search for gravitationally lensed quasars. 

In particular: 
\begin{itemize}
\item we have described each method, providing pre-selection criteria and assumptions,  and highlighting its strength and limitations; 
\item we have tested the methods on the recovery of known lenses to estimate the success rate of our suite of methods. We have used a list of ten known lenses (and NIQs) located in the current KiDS footprint (KiDS-DR3). 
Each method applied alone is able to recover only four out of ten of these lenses, but all three combined have reached, without any particular calibration, $70$\% of success rate. We adopt this number as the performance benchmark to {\sl catch them all} once the KiDS Survey is completed. Looser colour-cuts at pre-selection stage bring the recovery fraction to 80\%.
\item we have provided a list of good to excellent candidates recovered by one or more of our searching methods (Tables~\ref{tab:candidates_best} and \ref{tab:candidates_all}); 
\item we have confirmed the first of these high-graded new KiDS candidates (KiDS~1042+0023) in a spectroscopic run.
\item Finally we have discussed future prospects and ideas on how to improve the pre-selection criteria adopted in this paper. In particular we lists four different possibilities to minimize the number of lenses lost due to WISE infrared colours, without dramatically increase the number of contaminants and two possibilities to improve the optical pre-selection criteria. 
We will soon repeat the same analysis applying the methods with the improved pre-selection criteria described in Sec.~\ref{future} to the KiDS-DR4, which will cover about 900deg$^2$, and thus doubling the area, we will theoretically double the number of lensed quasar candidates. 
\end{itemize}

\begin{table*}
\caption{Candidates from KiDS-DR3 with grade $\geq2$ (from 1 to 4, with 4 being a sure lens).}
\label{tab:candidates_all}
\begin{center}
\begin{tabular}{lccccl}
\hline
{\bf ID}		&	{\bf RA (J2000)}	&	{\bf DEC (J2000)}	&	{\bf Methods} 	& {\bf Grade} 	& 	{\bf Notes} \\
\hline
KIDS0219-3430	&	02$:$19$:$36	&	-34$:$30$:$36 	& 	BaROQuES 			& 2.0			& \\
KIDS0237-3408	&	02$:$37$:$20	&	-34$:$08$:$24	&	DIA 				& 3.0			& \\		
KIDS0256-3101	&	02$:$56$:$08	&	-31$:$01$:$09 	& 	BaROQuES 			& 2.5			& \\
KIDS0313-3016	&	03$:$13$:$10 	& 	-30$:$16$:$26 	& 	BaROQuES 			& 3.0			& Very low separation between components $<0.2"$ \\
KIDS0324-3109	&	03$:$24$:$55	&	-31$:$09$:$18	&	DIA 				& 2.0			& \\
KIDS0834-0139	&	08$:$34$:$40 &	-01$:$39$:$08	&	DIA 				& 3.0			& Not a lens. TNG Spectroscopy reveals that the two objects are stars.\\
KIDS0835+0003	&	08$:$35$:$56	&	+00$:$03$:$06	&	DIA 				& 2.0			& \\
KIDS0838+0124	&	08$:$38$:$41	&	+01$:$24$:$5 	&	Multiplet 			& 2.0			& 	\\
KIDS0841-0008	&	08$:$41$:$01	&	-00$:$08$:$58	&	DIA 				& 2.0			& \\
KIDS0847-0013	&	08$:$47$:$10 	&	-00$:$13$:$03 	& 	DIA 				& 2.5			& One of the images has a QSO-like SDSS spectrum (z$=0.628$)	 \\
KIDS0848+0048	&	08$:$48$:$28	&	+00$:$48$:$34	&	DIA 				& 3.0			& \\	
KIDS0858-0147	&	08$:$58$:$34	&	-01$:$47$:$54 	& 	BaROQuES 			& 2.5	   & \\
KIDS0859+0159	&	08$:$59$:$03	&	+01$:$59$:$16 	& 	BaROQuES 			& 1.5			& Caution: it could be a QSO+gal system \\
KIDS0901+0111	&	09$:$01$:$03	&	+01$:$11$:$57	&	DIA 				& 3.0			& \\		
KIDS0904-0052	&	09$:$04$:$34	&	-00$:$53$:$30 	&	Multiplet 			& 2.0			& 	\\
KIDS0907+0003	& 	09$:$07$:$10	& 	+00$:$03$:$21 	& 	Multiplet			& 3.0	&   Possible NIQ \\
KIDS0914+0130	&	09$:$14$:$03	&	+01$:$30$:$34	&	DIA 				& 2.0			& \\
KIDS0916+0020	&	09$:$16$:$49	&	+00$:$20$:$47	&	DIA 				& 3.0			& \\		
KIDS0925+0021	&	09$:$25$:$40	&	+00$:$21$:$36 	&	Multiplet 			& 2.0			& 	\\
KIDS1028+0052	&	10$:$28$:$14	&	+00$:$52$:$56	&	Multiplet 			& 2.5 		 	& 	\\
KIDS1046+0017	&	10$:$46$:$56	&	+00$:$17$:$58	&	DIA 				& 2.5			& \\	
KIDS1114+0011	&	11$:$14$:$55	&	+00$:$11$:$14	&	DIA 				& 3.0 			& \\	
KIDS1133+0253	&	11$:$33$:$55	&	+02$:$53$:$42	& 	BaROQuES 			& 1.5			& Caution: f1/f2>>1 \\
KIDS1134+0139	&	11$:$34$:$37	&	+01$:$39$:$48	&	DIA 				& 3.0			& \\		
KIDS1138+0038	&	11$:$38$:$42	&	+00$:$38$:$27 	& 	BaROQuES 			& 2.0			& \\	
KIDS1139+0135	&	11$:$39$:$07	&	+01$:$35$:$56	& 	BaROQuES 			& 1.5			& Caution: it could be a QSO+gal system \\
KIDS1143-0118	&	11$:$43$:$50	&	-01$:$18$:$45	&	DIA 				& 2.5			& \\	
KIDS1143+0235	&	11$:$43$:$58	&	+02$:$35$:$08	&	DIA 				& 2.0			& \\
KIDS1146+0102	&	11$:$46$:$51	&	+01$:$02$:$40 	& 	BaROQuES 			& 2.0			& \\
KIDS1148+0218	&	11$:$48$:$31	&	+02$:$18$:$35 	& 	BaROQuES 			& 2.0			& \\
KIDS1152-0030	&	11$:$52$:$42	&	-00$:$30$:$13	&	DIA 				& 2.0		& Very faint \\
KIDS1201+0132	&	12$:$01$:$28	&	+01$:$32$:$56	&	DIA 				& 1.5			& \\
KIDS1202-0138	&	12$:$02$:$21	&	-01$:$38$:$08	&	Multiplet 			& 2.0			& 	\\
KIDS1209+0023	&	12$:$09$:$49	&	+00$:$23$:$59 	&	Multiplet 			& 2.0			& 	\\
KIDS1213+0255	&	12$:$13$:$52	&	+02$:$55$:$46	&	DIA 				& 2.0			& \\
KIDS1219+0251	&	12$:$19$:$50	&	+02$:$51$:$05 	& 	BaROQuES 			& 1.5			& \\
KIDS1220+0035	&	12$:$20$:$29	&	+00$:$35$:$26	&	Multiplet 			& 2.5			& 	\\
KIDS1224+0043	&	12$:$24$:$21	&	+00$:$43$:$30 	& 	BaROQuES 			& 1.5			& Both components have low brightness \\
KIDS1224+0135	&	12$:$24$:$48	&	+01$:$35$:$55	&	DIA 				& 2.0			& One of the components has an ETGs spectrum in SDSS. \\
KIDS1227+0004	&	12$:$27$:$36	&	+00$:$04$:$49	&	DIA 				& 2.0			& \\
KIDS1406-0112	&	14$:$06$:$22	&	-01$:$12$:$31	&	DIA 				& 2.5			& One of the images has a QSO-like SDSS spectrum (z=1.154) \\
KIDS1409+0256	&	14$:$09$:$18	&	+02$:$56$:$19 	& 	BaROQuES 			& 2.5			& \\
KIDS1412+0054	&	14$:$12$:$41	&	+00$:$54$:$59 	&	Multiplet 			& 2.0			& Separation bigger than 5$"$	\\
KIDS1414-0028	&	14$:$14$:$28	&	-00$:$28$:$08	&	DIA 				& 2.0			& Caution: could be galaxies. \\
KIDS1417-0032	&	14$:$17$:$27	&	+-00$:$32$:$28	&	DIA 				& 2.0			& \\
KIDS1420-0059	&	14$:$20$:$11	&	-00$:$59$:$11 	& 	BaROQuES 			& 2.5	& 	\\
KIDS1421+0157	&	14$:$21$:$31	&	+01$:$57$:$53	&	Multiplet 			& 2.5	& Possible Arc   \\
KIDS1429-0203	&	14$:$29$:$09	&	-02$:$03$:$03	&	DIA 				& 2.0			& \\
KIDS1435+0215	&	14$:$35$:$07	&	+02$:$15$:$56 	& 	BaROQuES 			& 2.0			& \\
KIDS1442-0147	&	14$:$42$:$14.	&	-01$:$47$:$40 	& 	BaROQuES 			& 2.5		& \\
KIDS1442-0126	&	14$:$42$:$18	&	+-01$:$26$:$24	&	BaROQuES 			& 2.0			& \\
KIDS1448-0216	&	14$:$48$:$40	&	-02$:$16$:$23 	& 	BaROQuES 			& 2.0			& \\
KIDS1449+0045	&	14$:$59$:$15	&	+00$:$45$:$07	&	BaROQuES 			& 2.0 & \\
KIDS1453+0039	&	14$:$53$:$47	&	+00$:$39$:$27	&	BaROQuES 			& 2.5		& One of the components has a QSO-like SDSS spectrum (z=1.093) \\
KIDS1459-0241	&	14$:$59$:$57	&	-02$:$41$:$46	&	BaROQuES 			& 2.0 & \\
KIDS1518+0056	&	15$:$18$:$42	&	+00$:$56$:$09	&	BaROQuES 			& 1.5		& Caution: the components seem to have different colours\\
KIDS1530-0034	&	15$:$30$:$02	&	-00$:$34$:$15 	& 	BaROQuES 			& 1.5			& \\
KIDS1544-0009	&	15$:$44$:$55	&	-00$:$09$:$34	&	BaROQuES 			& 1.5 			& \\
KIDS1547-0014	&	15$:$47$:$19	&	-00$:$14$:$21	&	BaROQuES 			& 1.5			& \\ 
KIDS1550+0221	&	15$:$50$:$57	&	+02$:$21$:$46	&	DIA 				& 2.5  & Caution: the components seem to have different colours\\
KIDS1551-0009	&	15$:$51$:$34	&	-00$:$09$:$02 	& 	BaROQuES 			& 1.5			& \\
\hline
\end{tabular}
  \end{center}
\end{table*}

\begin{table*}
\caption{...Continued from Table 3}
\begin{center}
\begin{tabular}{lccccl}
\hline
{\bf ID}		&	{\bf RA (J2000)}	&	{\bf DEC (J2000)}	&	{\bf Method} 	& 	{\bf Grade} & 	{\bf Notes} \\
\hline
KIDS2204-3222	&	22$:$04$:$22	&	-32$:$22$:$57 	& 	BaROQuES 		&	2.5			& \\
KIDS2219-3056	&	22$:$19$:$45	&	-30$:$56$:$01	&	BaROQuES 		&	2.0			& \\
KIDS2224-3154 	&	22$:$24$:$53 	&	-31$:$54$:$35 	& 	DIA 			&	2.5			& \\
KIDS2228-2837	&	22$:$28$:$46	&	-28$:$37$:$48 	& 	BaROQuES 		&	2.5			& \\
KIDS2232-3220	& 	22$:$32$:$07	& 	-32$:$20$:$22 	& 	BaROQuES 		&	3.0			& Very low separation between components $<0.2"$ \\	
KIDS2232-3417	&	22$:$32$:$29	&	-34$:$17$:$04 	&	Multiplet 		& 2.0 		& 	\\
KIDS2236-3400	&	22$:$36$:$56	&	-34$:$00$:$02	&	DBaROQuES		& 2.0			& \\
KIDS2241-3404	&	22$:$41$:$54	&	-34$:$04$:$42 	& 	BaROQuES 		& 2.0			& \\	
KIDS2243-3428	&	22$:$43$:$37 	&	-34$:$28$:$20 	& 	BaROQuES 		& 2.0			&  Possible QUAD, however colours are not convincing \\	
KIDS2259-3218	&	22$:$59$:$16	&	-32$:$18$:$41 	& 	BaROQuES 		& 2.0 			& Possible QUAD \\		
KIDS2300-3104	&	23$:$00$:$12	&	-31$:$04$:$07	&	DIA 			& 2.5		& \\	
KIDS2318-3222	&	23$:$18$:$06	&	-32$:$22$:$17 	& 	DIA 			& 2.5				& \\
KIDS2335-3042	&	23$:$35$:$49	&	-30$:$42$:$05 	& 	BaROQuES 		& 2.5				& \\
\hline
\end{tabular}
  \end{center}
\end{table*}

\section*{Acknowledgments} 
CS has received funding from the European Union's Horizon 2020 research and innovation programme under the Marie Sk\l odowska-Curie actions grant agreement No 664931. 
NRN acknowledges financial support from the European Union's Horizon 2020 research and innovation programme under the Marie Sk\l odowska-Curie grant agreement No 721463 to the SUNDIAL ITN network. 
CT, LVEK and GV are supported through a NWO VICI grant (Project No 639.043.308). 
MB is supported by the Netherlands Organization for Scientific Research, NWO, through grant No 614.001.451, and by the Polish National Science Center under the contract UMO-2012/07/D/ST9/02785.
KK acknowledges support by the Alexander von Humboldt Foundation. 
The authors wish to thank the anonymous referee for his/her constructive report that helped in improve the final version of the paper.

The project is based on data products from observations made with ESO Telescopes at the La Silla Paranal Observatory under programme IDs 177.A-3016, 177.A-3017 and 177.A-3018, and on data products produced by Target/OmegaCEN, INAF-OACN, INAF-OAPD and the KiDS production team, on behalf of the KiDS consortium. OmegaCEN and the KiDS production team acknowledge support by NOVA and NWO-M grants.

This publication makes also use of data products from the Two Micron All Sky Survey, which is a joint project of the University of Massachusetts and the Infrared Processing and Analysis Center/California Institute of Technology, funded by the National Aeronautics and Space Administration and the National Science Foundation. 
This publication makes use of data products from the Wide-field Infrared Survey Explorer, which is a joint project of the University of California, Los Angeles, and the Jet Propulsion Laboratory/California Institute of Technology, funded by the National Aeronautics and Space Administration. 
Finally, spectroscopic data presented in the paper have been obtained with the Italian Telescopio Nazionale Galileo (TNG) operated on the island of La Palma by the Fundación Galileo Galilei of the INAF (Istituto
Nazionale di Astrofisica) at the Spanish Observatorio del Roque de
los Muchachos of the Instituto de Astrofisica de Canarias.




\bsp	
\label{lastpage}
\end{document}